\documentclass[reprint,amsmath,amssymb,aps,floatfix]{revtex4-2}

\usepackage{graphicx}
\usepackage{dcolumn}
\usepackage{bm}
\usepackage[dvipsnames]{xcolor}
\usepackage{mathtools}
\usepackage{multirow}

\usepackage{hyperref}
\hypersetup{
    colorlinks=true,
    linkcolor=blue,
    citecolor=blue,
    urlcolor=blue
    }

\begin{document}



\title{In-depth Analysis of Anisotropic Magnetoconductance in Bi$_2$Se$_3$ thin films with electron-electron interaction corrections}

\author{Satyaki Sasmal}
\email[Corresponding author: ]{satyakisasmal@tifrh.res.in}
\affiliation{Tata Institute of Fundamental Research, Hyderabad, India}

\author{Joynarayan Mukherjee}
\affiliation{Tata Institute of Fundamental Research, Hyderabad, India}

\author{Dhavala Suri}
\affiliation{Tata Institute of Fundamental Research, Hyderabad, India}

\author{Karthik V. Raman}
\email[Corresponding author: ]{kvraman@tifrh.res.in}
\affiliation{Tata Institute of Fundamental Research, Hyderabad, India}


\begin{abstract}
A combination of out-of-plane and in-plane magnetoconductance (MC) study in topological insulators (TI) is often used as an experimental technique to probe weak anti-localization (WAL) response of the topological surface states (TSSs). However, in addition to the above WAL response, weak localization (WL) contribution from conducting bulk states are also known to coexist and contribute to the overall MC; a study that has so far received limited attention. In this article, we accurately extract the above WL contribution by systematically analyzing the temperature and magnetic field dependency of conductivity in Bi$_2$Se$_3$ films. For accurate analysis, we  quantify the contribution of electron-electron interactions to the measured MC which is often ignored in recent WAL studies. Moreover, we show that the WAL effect arising from the TSSs with finite penetration depth, for out-of-plane and in-plane magnetic field can together explain the anisotropic magnetoconductance (AMC) and, thus, the investigated AMC study can serve as a useful technique to probe the parameters like phase coherence length and penetration depth that characterise the TSSs in 3D TIs. We also demonstrate that increase in bulk-disorder, achieved by growing the films on amorphous SiO$_2$ substrate rather than on crystalline Al$_2$O$_3$(0001), can lead to stronger decoupling between the top and bottom surface states of the film.

\end{abstract}

\maketitle

\section{Introduction}

Three-dimensional topological insulators (3D-TIs) are a novel class of materials with conducting surface states within the bulk bandgap \cite{fu2007topological,hasan2010}. The conducting topological surface states (TSSs) are protected by the time reversal symmetry (TRS) \cite{zhang2009topological,moore2010birth} which make them robust against non-magnetic impurity scattering. In these surface states, the spin-momentum locking \cite{hasan2010,moore2010birth,qi2011} and the $\pi$ Berry phase \cite{ghaemi2010plane, lu2011competition} associated with the massless Dirac fermions collectively suppress backscattering of charge carriers giving rise to a weak antilocalization (WAL) response. As a result of these suppressed backscattering, 3D-TIs are extremely attractive material candidate for the development of energy efficient and robust electronic devices \cite{kong2011opportunities,he2019topological}. 

\par In experiments, the WAL behaviour is observed as an upward cusp near zero field in magnetoconductance (MC) measurements \cite{checkelsky2009quantum,chen2010gate,checkelsky2011bulk,he2011impurity,liu2012crossover,zhang2012interplay,yang2013,lang2013competing}.  WAL signal in the MC allows us to probe the electronic modifications in TSS due to finite film thickness \cite{lu2011competition,lang2013competing,lu2010massive}, surface-bulk coupling \cite{steinberg2011electrically,lin2013parallel}, inter-surface coupling \cite{shan2010effective,zhang2010crossover,lang2013competing}, magnetic doping \cite{liu2012crossover,zhang2012interplay} and magnetic proximity effect \cite{yang2013,li2017dirac,eremeev2013,mathimalar2020signature}. Recent theoretical \cite{tkachov2011weak} and experimental \cite{stephen2020weak, park2018disorder} investigations show that a WAL signal appears from the TSSs even when magnetic field is applied parallel to the film plane due to the finite penetration depth of the surface states. These observations suggest that the WAL signal from the TSSs, in both out-of-plane (OOP) and in-plane (IP) magnetic field, can contribute to the anisotropic magnetoconductance (AMC) of a TI. Also, apart from the WAL response of TSSs, weak localization (WL) response  \cite{lu2011weak,lu2014finite,zhang2012weak} of conducting bulk bands can also contribute to the overall MC.

\par Beside the MC study, analysis of temperature ($T$) dependent conductivity ($\sigma_{xx}$) can also provide information about the WAL signature of TSSs. With decreasing $T$, the WAL contribution is expected to show a logarithmic increase in $\sigma_{xx}$, but in systems like Bi$_2$Se$_3$, a logarithmic decrease is often observed \cite{liu2011electron,chen2011tunable,wang2011} due to the presence of electron-electron interaction (EEI) \cite{lee1985disordered, wang2011, chen2011tunable}. In TIs, the EEI is present within the bulk and at the surface \cite{lu2014finite} and the corresponding contribution to $\sigma_{xx}$ can be significant due to large g-factor \cite{wolos2016g,kohler1975g,analytis2010two,taskin2011berry}. Though EEI can also give a magnetic field ($B$) dependent correction to the $\sigma_{xx}$ with substantial magnitude \cite{wang2011,takagaki2012weak,liu2011electron, lu2014finite}, many of the WAL studies neglect this contribution of EEI.

\par In this article, we probe the combined effects of WAL, WL and EEI on the temperature response of $\sigma_{xx}$ \cite{lee1985disordered,chen2011tunable,takagaki2012weak,roy2013two,lu2014finite} by measuring,
\begin{equation}\label{slope}
    \kappa=\frac{\pi h}{e^2}\frac{\partial\sigma_{xx}}{\partial\ln{T}}
\end{equation}
which gives the slope of $\sigma_{xx}$ vs $\ln{T}$. When only EEI is present, $\kappa$ is expected to take a value between 0 and 1. Interestingly, in our experiments with Bi$_2$Se$_3$ thin films, we find $\kappa>1$ when WAL effect does not contribute to the $\sigma_{xx}$ (in high OOP magnetic fields). In Section \ref{subsec:EEI}, we resolve this discrepancy by considering the coexistence of WAL (surface) and WL (bulk) channels \cite{lu2014finite} in our samples. The overall $T$ dependency of $\sigma_{xx}$ results from the collective WAL, WL and EEI effect from these channels. Subsequently, in section \ref{subsec:MC} we analyse the OOP WAL response by considering the bulk states and TSSs with necessary EEI correction. We also analyze the IP WAL signal arising from the TSSs. In Section \ref{subsec:AMC}, we describe a formalism to analyse the measured AMC to characterize the TSSs. Additionally, we explore the effect of bulk disorder on WAL response by growing the film on different substrates viz. single crystalline Al$_2$O$_3$(0001) (c-Al$_2$O$_3$) and amorphous SiO$_2$. We show that bulk-disorder leads to a weakened coupling between the top and bottom TSSs supporting an enhanced contribution of TSS to the overall MC.

\section{Method}
Thin films of Bi$_2$Se$_3$ are prepared on c-Al$_2$O$_3$ and amorphous SiO$_2$ substrates using co-evaporation technique in a molecular beam epitaxy (MBE) system with a base pressure of $\sim 10^{-9}$ mBar. To minimize the presence of Se vacancies in Bi$_2$Se$_3$ films \cite{bansal2011epitaxial,bansal2014,chen2014molecular} excess Se flux (flux ratio of Bi:Se $\sim$ 1:15) is maintained during the growth process, followed by a capping layer of Se (5 nm) is deposited on top. The quality of Bi$_2$Se$_3$ films are analysed using Raman, in-situ RHEED and XRD. For transport measurements 16 nm Bi$_2$Se$_3$ films are mechanically patterned into Hall bars (channel width $\sim 500-1000\,\mu$m) under a microscope. Variable temperature insert cryostat with a base temperature of 1.5K has been used to perform low temperature magnetotransport measurements.

\section{Results and discussion}
\subsection{Film and device characterization}\label{subsec:growth}

\begin{figure}[t]
    \centering
    \includegraphics[width=\linewidth]{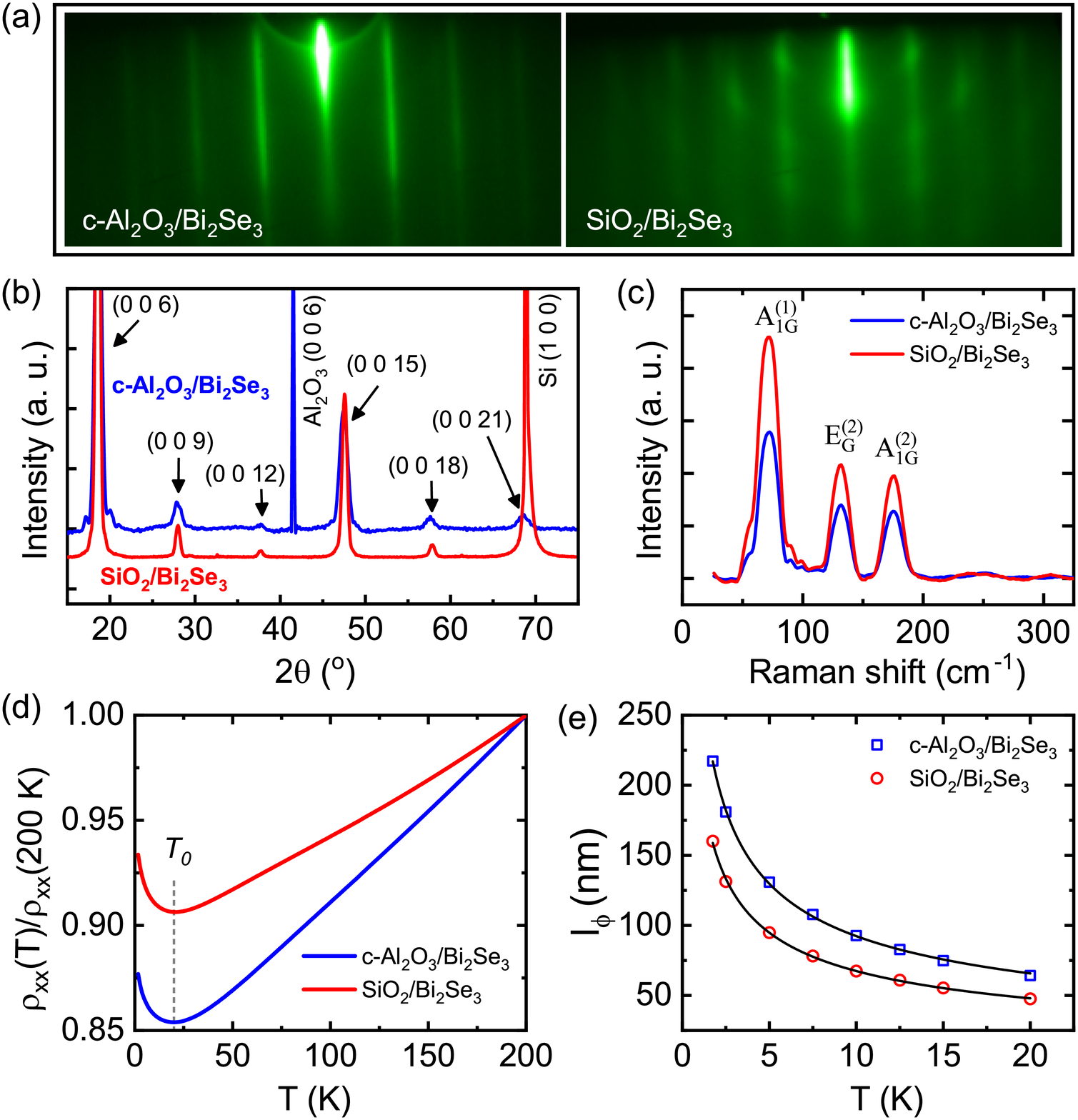}
    \caption{(a) RHEED image of Bi$_2$Se$_3$ films grown on c-Al$_2$O$_3$ (left) and SiO$_2$ (right). (b) XRD pattern of Bi$_2$Se$_3$ films on c-Al$_2$O$_3$ (blue) and SiO$_2$ (red) substrate. Peaks corresponding to Bi$_2$Se$_3$ planes are pointed out using arrows. (c) Raman spectra of Bi$_2$Se$_3$ films. Vibrational normal mode corresponding to the Raman peaks are labeled \cite{zhang2011raman}. (d) Normalized resistivity $\rho_{xx}/\rho_{xx}$(200 K) vs temperature $T$ for Bi$_2$Se$_3$ samples on both type of substrates. Dashed line indicates the temperature $T_0$ corresponding to lowest resistivity. Residual resistivity ratio (RRR), $\rho(200$ K$)/\rho(20$ K$)\sim1.17$ and $1.1$ for c-Al$_2$O$_3$/Bi$_2$Se$_3$ and SiO$_2$/Bi$_2$Se$_3$ films, respectively. (e) Phase coherence length $l_\phi$ as a function of $T$, measured from the magnetoconductance study at different temperatures. The solid black lines represents the best fit with $T^{-p/2}$.}
    \label{Fig: RHEED+R-T}
\end{figure}

The formation of Bi$_2$Se$_3$ phase in our films is confirmed by Raman study which shows the normal modes (figure \ref{Fig: RHEED+R-T}(c)) corresponding to the vibrations parallel ($A_{1g}^{1}$ and $A_{1g}^{2}$) and normal ($E_g^2$) to c-axis \cite{zhang2011raman} of Bi$_2$Se$_3$ crystal. Sharp vertical streaks in RHEED pattern (figure \ref{Fig: RHEED+R-T}(a), left) of Bi$_2$Se$_3$ films on c-Al$_2$O$_3$ imply atomically flat 2D surface with small domains. For Bi$_2$Se$_3$ films on SiO$_2$ substrate, streaks in RHEED pattern with modulated intensity (figure \ref{Fig: RHEED+R-T}(a), right) suggests flat surface with small multilevel steps. XRD $\theta-2\theta$  analysis, using Cu $K_\alpha$ source, shows (0 0 3) family of planes (figure \ref{Fig: RHEED+R-T}(b)) suggesting c-axis oriented growth of Bi$_2$Se$_3$ even on SiO$_2$ substrate \cite{jerng2013ordered}. In the case of SiO$_2$ substrate, in-plane amorphous growth is expected which was noted using isotropic RHEED pattern with in-plane sample rotation \cite{bansal2014}. Despite the above growth conditions in SiO$_2$/Bi$_2$Se$_3$ films, studies in literature have demonstrated the presence of robust Dirac surface states \cite{bansal2014,liu2015gate}. 

\par Hall measurement of the films grown on the two substrates reveal an electron carrier concentration, $n_H\sim2-3\times10^{13}$ cm$^{-2}$ and mobility, $\mu$ in the range of $100-200$ cm$^2$/V$\cdot$s. For all the batch of samples, comparatively, c-Al$_2$O$_3$/Bi$_2$Se$_3$ films show a higher $\mu$ than the SiO$_2$/Bi$_2$Se$_3$ films. While cooling down, both the films show a metallic signature with an upturn in resistivity at lower temperatures (figure \ref{Fig: RHEED+R-T}(d)). Here, SiO$_2$/Bi$_2$Se$_3$ films show relatively gradual decrease in the resistivity  suggesting the presence of higher bulk disorder in comparison to the films on c-Al$_2$O$_3$. However, in both the samples, the upturn in resistance appears at almost the same temperature $T_0\sim 20$ K indicating that the underlying mechanism to the upturn is independent of choice of substrate. We discuss this in detail in the following section.

\subsection{Evidence of EEI and background correction}\label{subsec:EEI}

The observed resistivity upturn in figure \ref{Fig: RHEED+R-T}(d) could originate from different mechanisms such as disorder \cite{brahlek2014,brahlek2015}, WL \cite{hikami1980} and EEI \cite{lee1985disordered}. In highly disordered metal, an insulating transition emerge if the mean free path of the electrons ($l_e$) becomes smaller than the Fermi wavelength ($1/k_F$). Based on the above argument, Ioffe and Regel derived \cite{ioffe1960} a critical limit for the mobility, $\mu_{IR}\approx (e/\hbar)/(3\pi^2 n_H/d)^{2/3}$ ($d$ is the thickness of the film), below which a disorder driven insulating transition can be observed. Mobility in our 16 nm Bi$_2$Se$_3$ films (obtained from Hall measurements) is found to be much above the critical limit ($\mu_{IR}\sim 19$ cm$^2$/V$\cdot$s, see section 1 in supplementary information for more details), allowing us to rule out the strong upturn response in resistivity due to film disorder effects. Additionally, the observation of similar resistivity upturn at $\sim 20$ K in both the samples also confirm these findings.

\begin{figure}[t]
    \centering
    \includegraphics[width=\linewidth]{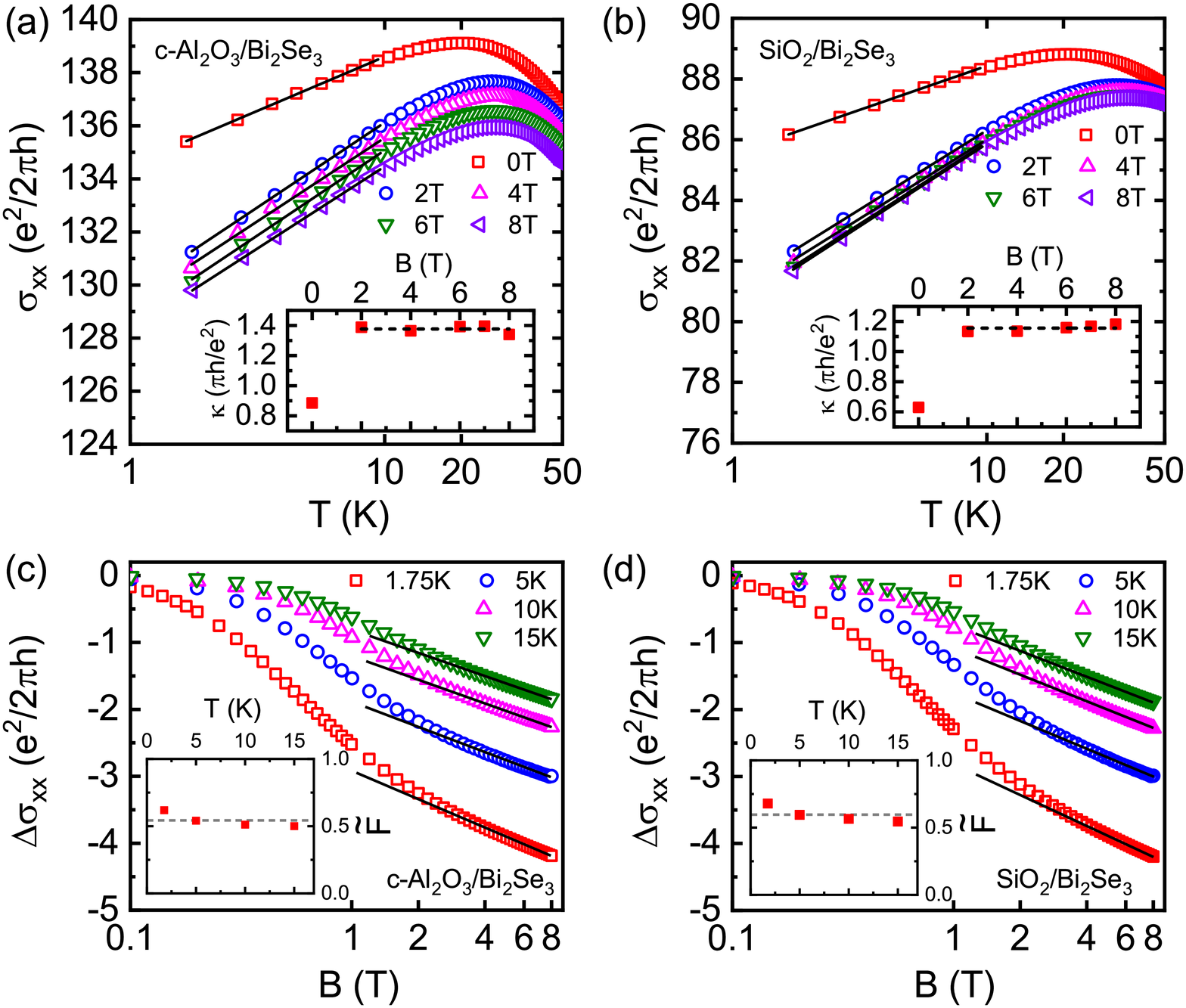}
    \caption{Conductance $\sigma_{xx}$ vs temperature $T$ for Bi$_2$Se$_3$ film on (a) c-Al$_2$O$_3$ and (b) SiO$_2$ with different strength of magnetic field $B$, normal to the film plane. Inset shows the measured value of slope $\kappa$ from the $\ln{T}$ fit (solid black lines in main figure).
    (c) and (d) shows the $\Delta\sigma_{xx}=\sigma_{xx}(B)-\sigma_{xx}(0)$ vs $B$ (parallel to the film plane) at different temperatures for c-Al$_2$O$_3$/Bi$_2$Se$_3$ and SiO$_2$/Bi$_2$Se$_3$ films respectively. In the inset the measured value of Coulomb screening factor $\tilde{F}$ from the $\ln{B}$ fit (solid black line in the main figure) is shown which is nearly independent with $T$. Dashed black line is a guide to the eye.}
    \label{Fig: GvsT}
\end{figure}

\par In a weakly disordered system where $l_e$ is larger than $1/k_F$, EEI accompanied by the impurity scattering can result in a reduction of electron density of states near Fermi level ($E_F$) \cite{altshuler1979zero}. For 3D, the temperature dependency due to EEI shows $\sigma_{xx}^{EEI}\propto \sqrt{T}$ \cite{lee1985disordered}. However, the effect becomes two dimensional (2D) when thermal diffusion length ($l_T=\sqrt{\hbar D/k_BT}$, $D$ is the diffusion constant) becomes larger than the film thickness and gives a $\ln{T}$ dependency \cite{lee1985disordered} as follows,
\begin{equation}\label{Eq:EEI_T}
    \Delta\sigma_{xx}^{EEI}=\frac{e^2}{\pi h}(1-\frac{3}{4}\tilde{F})\ln(\frac{T}{T_{EEI}})
\end{equation}
here, $\tilde{F}$ is Coulomb screening factor and $T_{EEI}$ is the characteristic temperature for EEI effect. We have measured $l_T\sim 150$ nm in 16 nm Bi$_2$Se$_3$ films which suggest the 2D nature of the EEI in our samples.

Beside the effect of EEI, in a system with strong spin-orbit interaction, the quantum interference (QI) effects i,e WAL or WL behaviour of the charge carriers also produces an increase or decrease in conductivity with decreasing temperature \cite{hikami1980},
\begin{equation}\label{WL_T}
    \Delta\sigma_{xx}^{QI}=\frac{e^2}{2\pi h}\tilde{A}p\ln(\frac{T}{T_{QI}})
\end{equation}
where $\tilde{A}=+1$ and $-1$, respectively for WL and WAL effect. $T_{QI}$ is the characteristic temperature for the QI effects. $p$ can be defined by the $T$ dependency of phase coherence length, $l_\phi\propto T^{-p/2}$ and it depends on the dimentionality and type of dephasing mechanism present in the system \cite{lin2002recent}. For EEI in an effectively 2D system, $p=1$.

To understand the type of interaction responsible for the dephasing mechanism in our system, we have measured $l_\phi$ as a function of $T$ from the MC measurements with OOP magnetic field at different temperatures. $l_\phi$ is determined from the fitting of Hikami-Larkin-Nagaoka equation \cite{hikami1980},
\begin{equation}\label{Eq:HLN}
    \Delta \sigma_{xx}^{QI}=\frac{e^2}{2\pi h}\tilde{A}\left[ \psi\left(\frac{1}{2}+\frac{B_\phi}{B}\right)-\ln\left(\frac{B_\phi}{B}\right)\right]
\end{equation}
to the OOP WAL signal (shown in supplementary figure 2(a) and 2b). Here $\psi$ is the digamma function and the dephasing field $B_\phi= \hbar/4el_{\phi}^{2}$. $|\tilde{A}|$ gives the effective number of 2D channels contributing to the WAL effect. Ideally, in 3D TI films, the presence of two TSSs (top and bottom) should give $|\tilde{A}|=2$. But, in case of Bi$_2$Se$_3$, $E_F$ lies inside the bulk conduction band (CB) due to high concentration of Se vacancies and the conducting bulk states weakly couple the top and bottom TSSs \cite{bansal2012thickness,chen2010gate,kim2011thickness} making  $|\tilde{A}|$ smaller than 2. From the fitting of equation (\ref{Eq:HLN}), measured value of $\tilde{A}$ up to $T=20$ K, is observed to be nearly independent of temperature (shown in supplementary figure 2(c)) with average value $|\tilde{A}|\sim1.2$ and $1.5$ for c-Al$_2$O$_3$/Bi$_2$Se$_3$ and SiO$_2$/Bi$_2$Se$_3$ films, respectively. In figure \ref{Fig: RHEED+R-T}(e) we plot $l_\phi$ vs $T$ and the fitting shows $p\sim0.98$ for both type of films confirming the presence of EEI in our system.

Since, in our Bi$_2$Se$_3$ films, WAL behaviour (confirmed by MC study) and EEI (confirmed by measured $p\sim1$) coexists, both of them contribute to the $\ln{T}$ behaviour of $\sigma_{xx}$ as given by equation (\ref{Eq:EEI_T}) and equation (\ref{WL_T}). At $B=0$, the overall slope (equation (\ref{slope})) can be written as $\kappa=\frac{1}{2}\tilde{A}p + (1-\frac{3}{4}\tilde{F})$. Here, the contribution of WAL or WL effect to the temperature dependence of $\sigma_{xx}$ can be suppressed by applying high magnetic fields \cite{chen2011tunable, lee1985disordered} and the value of the screening factor $\tilde{F}$ can be determined by measuring $\kappa$. In figure \ref{Fig: GvsT}(a) \& \ref{Fig: GvsT}(b), the logarithmic nature of $\sigma_{xx}$ with $T$ is shown for different magnitude of OOP magnetic fields. The inset shows the value of $\kappa$ for different $B$. For sufficiently large value of $B$ ($B>>B_\phi$), $\kappa$ saturates suggesting the complete suppression of WAL nature of charge carriers. The difference $\Delta\kappa=\kappa(B>>B_\phi)-\kappa(B=0)\approx-\frac{1}{2}\tilde{A}p$ is $+$ve i,e the effective $\tilde{A}<0$ which implies, WAL is a dominant mechanism in our films for small magnetic fields and the surface electrons are the majority charge carriers \cite{lu2014finite,lu2011weak}. However, in the above analysis, we have observed two discrepancies: (i) our measurement shows $\kappa > 1$ at large $B$ i,e $(1-\frac{3}{4}\tilde{F})>1$; this requires $\tilde{F}<1$ which is not physical as by definition, $0 \leq \tilde{F}\leq 1$ and (ii) from $\sigma_{xx}$ vs $T$ plot, we have measured $\Delta\kappa\sim0.491$ and $\sim 0.523$ for c-Al$_2$O$_3$/Bi$_2$Se$_3$ and
SiO$_2$/Bi$_2$Se$_3$ films, respectively. $\Delta\kappa$ provides the values of $|\tilde{A}|\sim1$ for both type of the films ($p\sim0.98$), which deviates from the measured $|\tilde{A}|$ ($\sim1.2$ and $\sim1.5$ for films on c-Al$_2$O$_3$ and
SiO$_2$, respectively) from HLN fitting of measured MC.

Similar discrepancy in the value of $\kappa$ was observed in Cu doped Bi$_2$Se$_3$ films \cite{takagaki2012weak} and Bi$_2$Te$_3$ films \cite{roy2013two} and attributed to strong spin orbit coupling (SOC) present in the TIs. Recent theoretical study \cite{lu2014finite} suggests that these discrepancies arise due to the presence of band-edge bulk channels which contribute to the WL effect beside the WAL effect from the TSSs. In the section \textbf{II.C} we have analysed low field MC considering the WAL effect from the surface states and WL effect from the band-edge bulk states and show that this is indeed the case in our Bi$_2$Se$_3$ samples.

Since in our samples the value of $\tilde{F}$ cannot be obtained by measuring $\Delta\kappa$, we have taken a different approach \cite{liu2011electron}. In the 2D limit, by neglecting SOC effects, the correction to MC ($\Delta\sigma_{xx}$) due to EEI can be written as \cite{lee1985disordered},

\begin{subequations}\label{Eq:ee_B}
\begin{gather}
    \Delta\sigma_{xx}^{EEI} (B)=-\frac{e^2}{2\pi h}\tilde{F} g_2(T,B)\\
    g_2 (T,B)=\int_{0}^{\infty}dt\ln\left\lvert{1-\left(\frac{h}{t}\right)^2}\right\rvert\frac{d^2}{dt^2}\left(\frac{t}{e^{t}-1}\right)
\end{gather}
\end{subequations}
here $h=g\mu_BB/k_BT$. The above equation is found to provide a decent approximation of EEIs in TIs \cite{wang2011,liu2011electron,chen2011tunable}. At high magnetic fields when Zeeman energy is much larger than the thermal activation energy ($h\gg1$), $g_2\approx \ln({h/1.3}$) and $\Delta\sigma_{xx}$, arising from EEI, shows a linear behaviour with $\ln{B}$. At $T=2$K, this limit requires $B\gg1.5$kOe. In our MC measurements, this linear response is observed  (figure \ref{Fig: GvsT}(c) \& \ref{Fig: GvsT}(d)) for $B>3$ T (along the film plane). At this high in-plane magnetic field, WAL signal from the TSSs can be neglected and from the linear fit (solid black line in figure \ref{Fig: GvsT}(c) \& \ref{Fig: GvsT}(d)) the effective value of $\tilde{F}$ is estimated \cite{liu2011electron} to be $\sim0.543$ and $\sim0.596$ for films on c-Al$_2$O$_3$ and SiO$_2$, respectively.

As shown in equation (\ref{Eq:ee_B}), we need the value of $g$ to estimate the contribution of EEI in $\sigma_{xx}$. Kohler and Wuchner have estimated the value of conduction electrons' $g$-factor to be $32$ and $23$ respectively for magnetic field's direction parallel and perpendicular to c-axis of Bi$_2$Se$_3$ crystal using Shubnikov-de Haas oscillation measurements \cite{kohler1975g}. Recently, more precise electron spin resonance (ESR) measurements by Wolos \textit{et al.} have shown the value of $g$-factor for conduction electron to be equal to $27.3\pm0.15$ and $19.48\pm0.07$ for magnetic field parallel and perpendicular to c-axis \cite{wolos2016g}. In the following analysis, the extracted value of $\tilde{F}$ and $g$-factor values, from ESR measurements, are used to determining the field dependence of  $\Delta\sigma_{xx}^{EEI}$ (equation (\ref{Eq:ee_B})) and subsequently the EEI background correction to the measured MC is performed.


\subsection{OOP and IP MC studies}\label{subsec:MC}

Studies by H. T. He \textit{et al.} \cite{he2011impurity} suggest the existence of magnetoresistance proportional to $B^2$ due to Lorentz deflection of bulk charge carriers. The corresponding correction to the MC can be written as, $\Delta G_{xx} (B) = 1/R(B) - 1/R(0)$, where according to Kohler's rule \cite{he2011impurity,olsen1962electron} $R(B)/R(0)=1+(\mu B)^2$, $\mu$ is the film mobility. However, due to relatively low $\mu$ of our films this contribution is very small as compared to the measured MC (see supplementary figure 3(a)) and therefore is neglected in our MC analysis.
\par In a Dirac system like topological insulators, the Berry phase acquired by the electrons after completing a closed loop on the Fermi surface, can be written as $\phi_b=\pi(1-\Delta/2E_F)$ \cite{lu2011competition,ghaemi2010plane}. Here, $\Delta$ is the Dirac mass and $E_F$ is the Fermi level referenced from the Dirac point. For the surface states, $\Delta\rightarrow0$, and the corresponding $\phi_b=\pi$ leads to a destructive quantum interference and thus the TSSs shows WAL effect. On the other hand, the bulk states have a finite gap $\Delta$ and when $E_F$ is near the conduction band-edge $\Delta/2E_F \rightarrow1$. In this situation, electrons in the gaped band-edge bulk states, goes through a constructive interference upon tracing a closed path, leading to WL effect. Therefore, the experimentally observed WAL in TIs could be a combined result of WL from band-edge bulk states and WAL from surface states \cite{lu2011weak,lu2014finite}.

\begin{figure}[t]
    \centering
    \includegraphics[width=\linewidth]{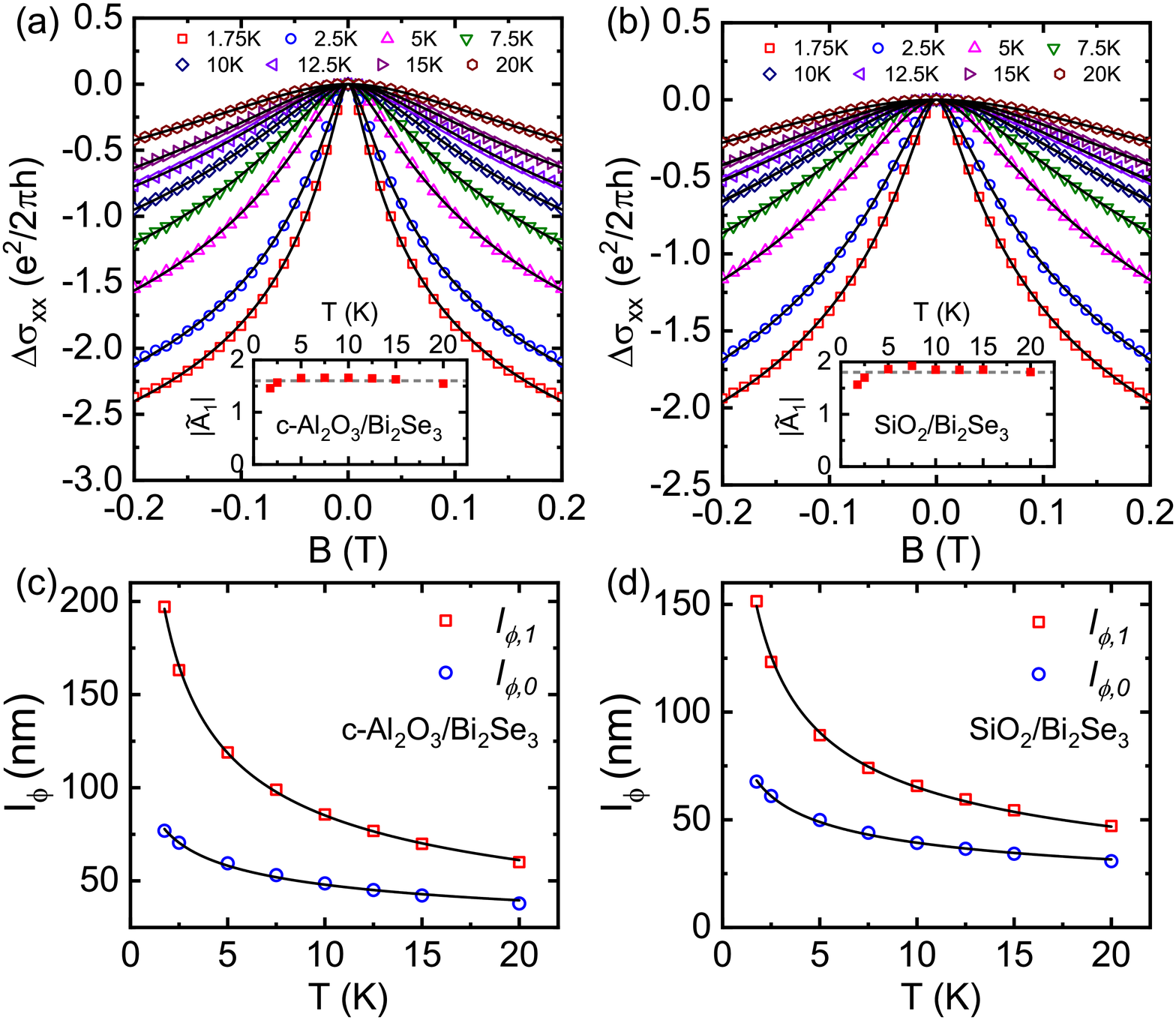}
    \caption{(a) and (b) shows the measured weak anti localization signal at perpendicular magnetic fields for c-Al$_2$O$_3$/Bi$_2$Se$_3$ and SiO$_2$/Bi$_2$Se$_3$ films respectively. Solid black lines show the fitting with equation (\ref{Eq:two_channel}). Extracted value of the prefactor $|\tilde{A}_1|$ is shown in the inset for the respective films. The dashed line (at $|\tilde{A}_1|=1.6$ and $1.8$ for c-Al$_2$O$_3$/Bi$_2$Se$_3$ and SiO$_2$/Bi$_2$Se$_3$) shows the average value of $|\tilde{A}_1|$.  (c) and (d) shows the estimated values of the phase coherence lengths ($l_{\phi,0}$ and $l_{\phi,1}$) for respectively c-Al$_2$O$_3$/Bi$_2$Se$_3$ and SiO$_2$/Bi$_2$Se$_3$ films. The solid black lines in (c) and (d) represents the fitting with $T^{-p/2}.$}
    \label{Fig:OOP-WAL}
\end{figure}

\begin{figure*}[]
    \centering
    \includegraphics[width=\textwidth]{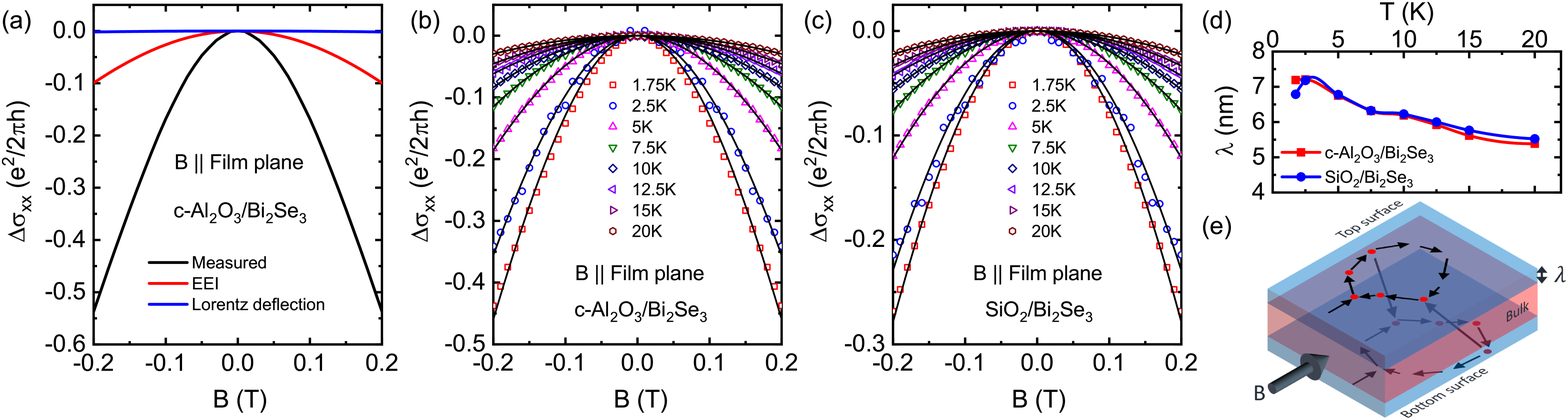}
    \caption{(a) Comparison of the measured in-plane weak anti localization signal (black line) with the estimated EEI correction (red line) for in-plane magnetic field and the correction due to Lorentz deflection of charge carriers (blue line). (b) and (c) shows the in-plane weak anti localization data at different temperatures after the contribution from electron-electron interaction is subtracted. Solid black line represents the best fit with equation (\ref{th}). (d) shows the estimated value of the penetration depth $\lambda$ from the fitting shown in (b) and (c). Solid lines are present to guide the eye. (e) A schematic showing the finite penetration depth of the surface states and the surface-surface scattering of charge carriers as a source of WAL signal when magnetic field is parallel to the film plane. }
    \label{Fig:IP-WAL}
\end{figure*}

In our Bi$_2$Se$_3$ films the Hall carrier density is $n_H\sim2-3\times10^{13}$ cm$^{-2}$ which places the $E_F$ near the bottom of the bulk conduction band \cite{kim2013coherent,liu2015gate}. However, capping the Bi$_2$Se$_3$ films with thick Se layer can reduce the top surface carrier concentration \cite{liu2015gate} significantly and can bring the surface $E_F$ closer to the Dirac point. This mismatch between the surface and bulk $E_F$ can further lead to a upward band bending near the top surface and can bring the bulk Fermi level closer to the conduction band-edge \cite{brahlek2015,mathimalar2020signature} making $\Delta/2E_F \rightarrow1$ for band-edge bulk states. Therefore, in our Bi$_2$Se$_3$ thin films, band-edge WL channel and surface WAL channels may coexist. The corresponding quantum correction formula for the MC in presence of OOP magnetic field can be written as \cite{lu2011weak},
\begin{equation}\label{Eq:two_channel}
    \Delta \sigma_{xx}^{QI}=\frac{e^2}{2\pi h} \sum_{i=0,1} \tilde{A_i}\left[ \psi\left(\frac{1}{2}+\frac{B_{\phi,i}}{B}\right) -\ln\left(\frac{B_{\phi,i}}{B}\right)  \right]
\end{equation}
where, $i=0$ corresponds to the $\Delta/2E_F\rightarrow1$ (WL) and $i=1$ corresponds to $\Delta/2E_F=0$ (WAL) situation. $B_{\phi,i}= \hbar/4el_{\phi,i}^{2}$. In the limit $\Delta/2E_F \rightarrow0$ and $\Delta/2E_F \rightarrow1$, $l_{\phi,i}$ represents the phase coherence length \cite{lu2011weak,lu2014finite}. To replicate more realistic scenario, we are assuming that the phase coherence length for surface states and band-edge bulk states are different i,e $l_{\phi,0}\neq l_{\phi,1}$. Theoretically $\tilde{A}_0=1$ and $\tilde{A}_1=-2$ (top and bottom 2D surface channel). In our analysis, we keep $\tilde{A}_1$ as a fitting parameter as the top and bottom surface states can be weakly coupled through the conducting bulk \cite{bansal2012thickness,chen2010gate,kim2011thickness}.

\par As mentioned in the previous section, the contribution of EEI needs to be removed from the MC data for accurate analysis. In figure \ref{Fig:OOP-WAL}(a) \& \ref{Fig:OOP-WAL}(b), we have shown the measured OOP MC of both films after the subtraction of EEI contributions. The fitting with equation \ref{Eq:two_channel} is shown by solid lines. Corresponding value of $\tilde{A}_1$  (shown in the inset) is observed to be almost independent of temperature. Dashed gray line shows the average value. The measured value of $l_{\phi,0}$ and $l_{\phi,1}$ is plotted with respect to $T$ (figure \ref{Fig:OOP-WAL}(c) \& \ref{Fig:OOP-WAL}(d)) and fitted with $T^{-p_i/2}$. We measure $p_1\sim0.95$ (corresponds to TSSs) for both types of films which agrees with the theoretically expected $p=1$ in case of 2D EEI. For the electrons in band-edge bulk states, we measure $p_0\sim0.55$ and $\sim0.63$ for c-Al$_2$O$_3$/Bi$_2$Se$_3$ and SiO$_2$/Bi$_2$Se$_3$ films respectively, which is smaller than the theoretical value. However, similar value of $p$ has been observed before in Bi$_2$Se$_3$ micro flakes \cite{chiu2013weak} and few-layer WTe$_2$ \cite{zhang2020crossover} in presence of EEI. The smaller value of $p_0$ may suggest the presence of other weak dephasing effects \cite{huang2007observation, lin2002recent} within the bulk which is beyond the scope of present work.


Our analysis shows $\tilde{A}_1\sim1.8$ for SiO$_2$/Bi$_2$Se$_3$ films which is higher than $\tilde{A}_1\sim1.6$ measured for c-Al$_2$O$_3$/Bi$_2$Se$_3$ films. This suggests comparably weaker coupling between top and bottom TSS in SiO$_2$/Bi$_2$Se$_3$ films. The phase coherence length of the TSSs ($l_{\phi,1}$) is relatively  bigger ($\sim33\,\%$ bigger at 1.75 K) in c-Al$_2$O$_3$/Bi$_2$Se$_3$ films whereas $l_{\phi,0}$ is marginally higher in these films. $l_\phi$ gives a measure of disorder present in the system and it gets smaller with increasing disorder \cite{kim2013coherent}. Comparably smaller value of $l_{\phi,i}$ in SiO$_2$/Bi$_2$Se$_3$ films may suggest the presence of higher disorder in these films which is also observed in the RHEED images and the value of $\mu$ obtained from the Hall measurements. The presence of disorder can effectively decouple the top and bottom surface states \cite{kim2013coherent,park2018disorder} resulting in a higher value of $\tilde{A}_1$ as observed in our SiO$_2$/Bi$_2$Se$_3$ samples. Previous reports have achieved similar decoupling between the top and bottom TSSs in TI thin films by tuning the Fermi level \cite{steinberg2011electrically, kim2013coherent}, modulating the film thickness \cite{brahlek2014} and controlling the disorder by varying the annealing conditions \cite{park2018disorder}.

From the extracted values of $\tilde{A}_i$ and $p_i$ we obtain $\Delta\kappa=-\frac{1}{2}\sum_i p_i\tilde{A}_i$ to be equal to 0.488 and 0.541 for c-Al$_2$O$_3$/Bi$_2$Se$_3$ and
SiO$_2$/Bi$_2$Se$_3$ films, respectively ($p_i$ corresponds to $T^{-p/2}$ fitting of $l_{\phi,i}$). These values are in very good agreement with the values measured from the $\sigma_{xx}$ vs $\ln{T}$ plot confirming the existence of two types of conducting channel in our samples. The presence of two types of conducting channels can also explain the measured $\kappa>1$ \cite{lu2014finite} in section \ref{subsec:EEI}. When multiple transport channels exist, $\kappa$ for $B\gg B_\phi$ can be written as \cite{tkavc2019influence}, $\kappa=\sum_j(1-\eta_j \tilde{F}_j)$, where $j$ runs over independent WL and WAL channels, $\eta_i$ is a scaling factor and $\tilde{F}_i$ is the Coulomb screening factor. $\eta$ has the values $3/4$ and $1$ for $\Delta/2E_F=0$ and 1, respectively. In comparison to the previous approach in Section \ref{subsec:EEI} to determine the overall value of $\tilde{F}$ using equation (\ref{Eq:ee_B}), the present expression of $\kappa$ includes weighted values of $\tilde{F}$ for the two independent transport channels.

\par Beside the usual OOP WAL response, we have also observed a strong WAL signal when field is applied parallel to Bi$_2$Se$_3$ film (solid black line in figure \ref{Fig:IP-WAL}(a)). A -ve MC in IP magnetic field can arise from the Lorentz deflection of charge carriers in the bulk of the TI \cite{he2011impurity}. EEI can also contribute to the IP MC as described in the previous section. Estimated value of these two effects is compared with the observed IP WAL signal in figure \ref{Fig:IP-WAL}(a). 

At low fields, the contribution from Lorentz deflection (\ref{Fig:IP-WAL}(a), solid blue line) is negligible as compared to the observed signal due to small $\mu$ of our Bi$_2$Se$_3$ samples. Estimated correction to the MC from EEI effect is also much smaller than the measured value (\ref{Fig:IP-WAL}(a), solid red line). This suggests the presence of nontrivial effects which contribute to strong IP WAL signal in our Bi$_2$Se$_3$ samples. Recent theoretical developments shows that TIs can exhibit IP MC due to the exponential decay of TSSs wavefunction into the bulk \cite{tkachov2011weak} which effectively results in a finite thickness of the 2D-TSSs. First principle studies of Bi$_2$Se$_3$ system \cite{zhang2010first} have estimated the penetration depth of the TSSs to be up to 2-3 quintuple layers (1 quintuple layer $\sim$ 0.955 nm \cite{zhang2011raman}). When magnetic field is applied along the film plane, magnetic flux through the finite cross-section of the TSSs (figure \ref{Fig:IP-WAL}(e)) can give rise to a -ve MC. Corresponding correction to the MC is derived by Tkachov and Hankiewicz \cite{tkachov2011weak}(TH) model,
\begin{equation}\label{th}
    \Delta \sigma_{xx} = \frac{e^2}{2\pi h}\tilde{A}_1\ln \left(1+\frac{B^2}{B_{||}^2} \right)
\end{equation}
where, $B_{||} = \sqrt{2\hbar B_{\phi,1}/(e\lambda^2)}$ and $\lambda$ is the penetration depth. For perfectly two dimensional TSS (i,e $\lambda = 0$) this correction vanishes.

figure \ref{Fig:IP-WAL}(b) \& \ref{Fig:IP-WAL}(c) shows the measured value of IP WAL signal at different temperatures after removing the EEI background. Correction due to Lorentz deflection of charge carriers has been neglected due to its negligible magnitude. Solid black lines represent the best fit using equation (\ref{th}) with $B_\parallel$ as the fitting parameter. The measured value of $\tilde{A}_1$ from figure \ref{Fig:OOP-WAL} is used in the fitting to consider the weak coupling between top and bottom surface states. The range of magnetic field used in the fitting is much greater than the dephasing field $B_{\phi,1}$ of the TSSs. Measured $B_{\phi,1}$ from the fitting of equation (\ref{Eq:two_channel}) (as shown in figure \ref{Fig:OOP-WAL}(a) \& \ref{Fig:OOP-WAL}(b)) has been used to compute the value of $\lambda$ using the TH model fitting. Our analysis shows similar value of $\lambda$ for the two types of films with $\lambda\sim7$ nm at 1.75 K. Our Bi$_2$Se$_3$ samples show much smaller value of $\lambda$ as compared to the recently reported $\lambda\sim50$ nm on 50 nm thick Te doped Bi$_2$Se$_3$ films \cite{stephen2020weak} that also introduces considerable structural disorder. H Park \textit{et al.} measured $\lambda\sim8.5$ nm for a $\sim$14 nm thick doped Bi$_2$Se$_3$ films. The measured $\lambda$ in our samples is higher than the previously estimated values from first principle calculations \cite{zhang2010first}. The reason behind large $\lambda$ could be the phase coherent trajectories of surface charge carriers through the conducting bulk \cite{lin2013parallel}, in a perpendicular direction to the film plane, coupling the top and bottom surface states weakly (as shown schematically in figure \ref{Fig:IP-WAL}(e)). However, we measure $2\lambda<d$ for all the temperatures suggesting a better decoupling between the TSSs which also supports the observation of $|\tilde{A}_1|$ close to the value 2. Interestingly, SiO$_2$/Bi$_2$Se$_3$ samples have a similar penetration depth and stronger decoupling (larger $|\tilde{A}_1|$) between the two TSSs as compared to c-Al$_2$O$_3$/Bi$_2$Se$_3$ films.

\subsection{AMC study}\label{subsec:AMC}

Since our observations show the effect of TSSs on the MC for both IP and OOP magnetic field originating from two distinct mechanisms, we now try to probe their combined response on the AMC of Bi$_2$Se$_3$ films. Recently, Hui Li \textit{et al.} \cite{li2019quantitative} studied the AMC of BiSbTeSe$_2$ thin films, but in their analysis, the WAL contribution of the 2D surface states at IP magnetic field has been neglected. In contrast, our study has shown a considerable IP WAL signal arising from the finite penetration depth of the surface states. 
\begin{figure*}[]
    \centering
    \includegraphics[width=\linewidth]{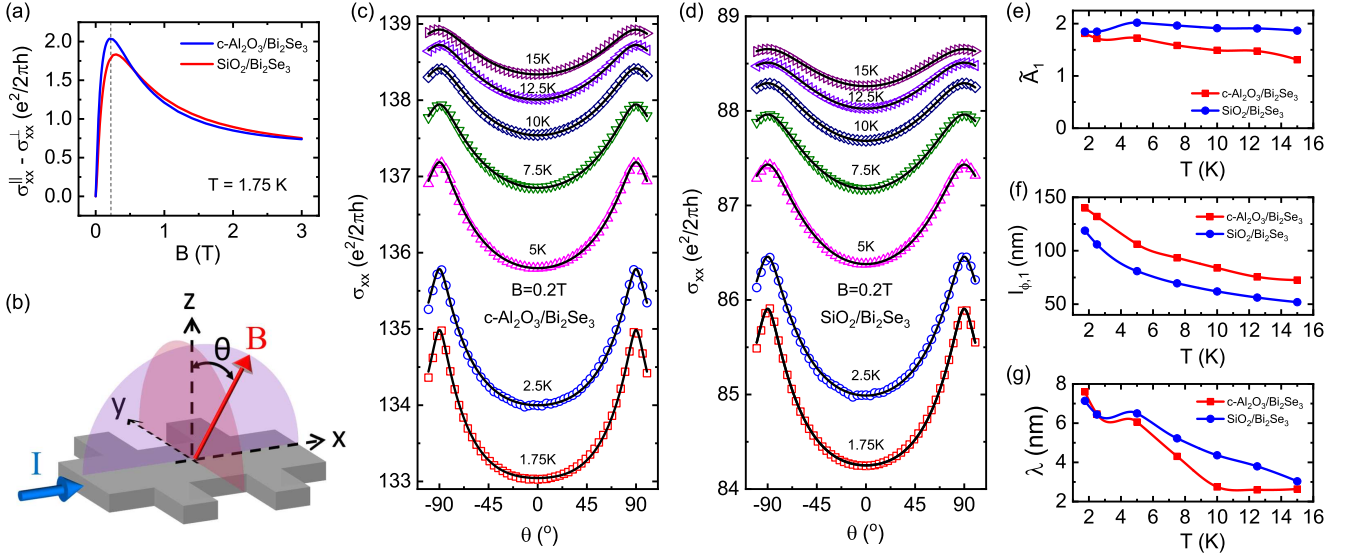}
    \caption{(a) Difference in the magnitude of conductance ($\sigma_{xx}^\parallel - \sigma_{xx}^\perp$) for in-plane ($\sigma_{xx}^{\parallel}$) and out-of-plane ($\sigma_{xx}^{\perp}$) configuration vs magnetic field strength. Dashed line roughly indicates the maxima of $\sigma_{xx}^\parallel - \sigma_{xx}^\perp$. (b) Schematic showing the rotation scheme of the magnetic field direction. Tilt angle $\theta$ is measured with respect to the normal to the film plane. (c) \& (d) Anisotropic magnetoconductance of c-Al$_2$O$_3$/Bi$_2$Se$_3$ and SiO$_2$/Bi$_2$Se$_3$ at magnetic field $B=0.2$ T and for temperatures up to 15 K. Solid black lines show the fitting with equation (\ref{Eq:HLN+TH}). (e), (f) and (g) show the measured values of fitting  parameters, $\tilde{A}_1$, $l_{\phi,1}$ and $\lambda$ respectively, at different temperatures.}
    \label{Fig:AMC}
\end{figure*}
\par At low temperatures where the effects of TSSs is dominant (for both IP and OOP configuration), we write an expression for the AMC by only considering the WAL behaviour for IP (equation (\ref{th})) and OOP (equation (\ref{Eq:two_channel})) magnetic field,
\begin{multline}\label{Eq:HLN+TH}
    \sigma_{xx}(\theta)=\sigma_{xx}(B=0)\\
    +\frac{e^2}{2\pi h}\tilde{A}_0\Bigg\{\psi\left(\frac{1}{2}+\frac{B_{\phi,0}}{|B\,cos\,\theta|}\right) -\ln\left(\frac{B_{\phi,0}}{|B\,cos\,\theta|}\right) \Bigg\}\\
    +\frac{e^2}{2\pi h}\tilde{A}_1\Bigg\{\ln \left[1+\left(\frac{B\,sin\,\theta}{B_{||}}\right)^2 \right]\\
     + \psi\left(\frac{1}{2}+\frac{B_{\phi,1}}{|B\,cos\,\theta|}\right) -\ln\left(\frac{B_{\phi,1}}{|B\,cos\,\theta|}\right) \Bigg\}
\end{multline}
Here, $\sigma_{xx}(B=0)$ is the value of the conductance when no magnetic field is applied. 2$^{nd}$ term describes the correction arising from the bulk band-edge WL channels due to the OOP component of the magnetic field. 3$^{rd}$ term represents the WAL effect from TSSs for IP and OOP component of $B$. $\theta$ is the angle between $B$ and the film-normal (see figure  \ref{Fig:AMC}(b)).

\par For our devices, the difference of IP and OOP MC is maximum around $B=0.2$ T (figure \ref{Fig:AMC}(a)). Keeping this fact into consideration, we have chosen to measure the anisotropy in MC with $B=0.2$ T. The rotation schemes of the measurements is shown in figure \ref{Fig:AMC}(b). Magnetic field is rotated in x-z plane and y-z plane. Our measurements show a highly anisotropic nature of MC with a sharp peak at $\theta=\pm 90^\circ$. We have observed no significant difference in the AMC for the rotation of $B$ in the x-z and y-z plane (see supplementary figure 4) which suggest the absence of IP anisotropy in the MC of our Bi$_2$Se$_3$ films. 

\par An OOP anisotropy in $\sigma_{xx}$ can appear from EEI background due to the anisotropy in the value of $g$-factor of Bi$_2$Se$_3$ \cite{wolos2016g}. For this purpose, the $g$-factor in equation (\ref{Eq:ee_B}) is replaced with $g=\sqrt{g_\perp^2 cos^2 \, \theta + g_\parallel^2 sin^2 \, \theta}$, where $g_\perp=27.3$ and $g_\parallel=19.48$ are the $g$-factor value of conduction electrons for OOP (parallel to c-axis) and IP (perpendicular to c-axis) magnetic field, respectively. The additional anisotropy in EEI due to high SOC \cite{fukuyama1982effects,markiewicz1984localization} is expected to be weak in our AMC study as these measurements were performed at low magnetic fields.

\par The measured AMC at different temperatures, after the EEI background correction using equation (\ref{Eq:ee_B}), is shown in figure \ref{Fig:AMC}(c) \& \ref{Fig:AMC}(d). The observed AMC is well fitted with equation (\ref{Eq:HLN+TH}) (solid black lines in figure \ref{Fig:AMC}(c) \& \ref{Fig:AMC}(d)). To reduce the number of fitting parameters and to keep the analysis simple, we have kept the WL correction fixed by keeping $\tilde{A}_0$ and $B_{\phi,0}$ fixed at the values, obtained in Section \ref{subsec:MC} from WAL analysis. The parameters $\tilde{A}_1$, $B_{\phi,1}$ and $B_\parallel$ in equation (\ref{Eq:HLN+TH}) are used as the fitting parameters in the analysis and $\sigma_{xx}(B=0)$ is measured experimentally.

\par The observed sharp peak in the AMC near $\theta=\pm90^\circ$ at low temperatures arises due to the WAL effect in TSSs for OOP component of the magnetic field. The WAL signal due to the finite thickness of TSSs for IP component of $B$ gives a correction to the peak height at $\theta=\pm90^\circ$. Fitting results of equation (\ref{Eq:HLN+TH}) shown in figure \ref{Fig:AMC}(e)-\ref{Fig:AMC}(g) yields similar value of $\tilde{A}$, $l_\phi$ and $\lambda$  to those, measured independently from OOP and IP WAL analysis, earlier in the section \ref{subsec:MC}. At 1.75 K, AMC study gives $l_{\phi,1}\sim 140$ nm and $\sim 118$ nm for c-Al$_2$O$_3$/Bi$_2$Se$_3$ and SiO$_2$/Bi$_2$Se$_3$ films respectively which are smaller than the values ($\sim 197$ nm and  $\sim 150$ nm, respectively) obtained from WAL study in section \ref{subsec:MC}. However, this mismatch in $l_{\phi,1}$ reduces with increasing temperature. Extracted value of $\tilde{A}_1$ from the fitting of equation (\ref{Eq:HLN+TH}) matches reasonably with the values measured from WAL study. Although the values of $\lambda$ obtained from AMC study (figure \ref{Fig:AMC}g) and IP WAL study (figure \ref{Fig:IP-WAL}d) match near lowest temperatures, it deviates considerably at higher temperatures showing almost half of the value obtained from IP WAL studies. This mismatch is possibly arising because the IP-WAL signal becomes significantly weaker at higher $T$ (See figure \ref{Fig:IP-WAL}(b) \& \ref{Fig:IP-WAL}(c)), possibly making the anisotropy in the magnetoconductance of the bulk Bi$_2$Se$_3$ comparable which has been neglected in our analysis. Nevertheless, equation (\ref{Eq:HLN+TH}) provides a good estimate of $\tilde{A}$, $l_\phi$ and $\lambda$ which are extracted simultaneously and explains the AMC data reasonably well. Our above observations indicate that the investigated AMC analysis can serve as a useful technique to probe the TSSs in 3D TIs.

\section{Conclusion}

In conclusion, our study of Bi$_2$Se$_3$ thin films show that EEIs give a considerable contribution to the total MC. We provide details of a systematic approach to remove these contributions for the analysis of WAL (for both IP and OOP configuration) and AMC. We show that the discrepancy in the $T$ and $B$ dependence of $\sigma_{xx}$ can be resolved by considering WL effects from the band-edge bulk states.  For IP magnetic fields, a strong WAL signal with nontrivial origin is observed that finds applicability of the TH model to explain our IP MC study. Additionally, we also convey the significance of finite penetration depth of TSSs in analyzing the AMC experiments. The fitting routine using a combination of HLN model and TH model explains the AMC in our Bi$_2$Se$_3$ films and the fitting parameters, which characterise the TSSs, show the expected trend. Importantly, we find that Bi$_2$Se$_3$ films on SiO$_2$ with stronger bulk-disorder than on c-Al$_2$O$_3$ can lead to a better decoupling of the two TSSs.

\section{Acknowledgement}
Authors thank Dr. T. N. Narayanan for providing access to Raman facility. We are also grateful to IIT Hyderabad for giving access to their XRD facility. Authors acknowledge funding support of the Department of Atomic Energy, Government of India, under Project Identification No. RTI 4007 and Science and Engineering Research Board Grant No. CRG/2019/003810.

\section*{Data Availability}
Data that support the findings of this study is available from the corresponding authors upon reasonable request. 


\begin{thebibliography}{10}
\expandafter\ifx\csname url\endcsname\relax
  \def\url#1{{\tt #1}}\fi
\expandafter\ifx\csname urlprefix\endcsname\relax\def\urlprefix{URL }\fi
\providecommand{\eprint}[2][]{\url{#2}}

\bibitem{fu2007topological}
Fu L, Kane C~L and Mele E~J 2007 {\em Phys. Rev. Lett.\/} {\bf 98} 106803

\bibitem{hasan2010}
Hasan M~Z and Kane C~L 2010 {\em Rev. Mod. Phys.\/} {\bf 82} 3045--3067

\bibitem{zhang2009topological}
Zhang H, Liu C~X, Qi X~L, Dai X, Fang Z and Zhang S~C 2009 {\em Nat. Phys.\/}
  {\bf 5} 438--442

\bibitem{moore2010birth}
Moore J~E 2010 {\em Nature\/} {\bf 464} 194--198

\bibitem{qi2011}
Qi X~L and Zhang S~C 2011 {\em Rev. Mod. Phys.\/} {\bf 83} 1057--1110

\bibitem{ando1998berry}
Ando T, Nakanishi T and Saito R 1998 {\em J. Phys. Soc. Japan\/} {\bf 67}
  2857--2862

\bibitem{suzuura2002crossover}
Suzuura H and Ando T 2002 {\em Phys. Rev. Lett.\/} {\bf 89} 266603

\bibitem{kong2011opportunities}
Kong D and Cui Y 2011 {\em Nat. Chem.\/} {\bf 3} 845--849

\bibitem{he2019topological}
He M, Sun H and He Q~L 2019 {\em Front Phys\/} {\bf 14} 43401

\bibitem{checkelsky2009quantum}
Checkelsky J~G, Hor Y~S, Liu M~H, Qu D~X, Cava R~J and Ong N 2009 {\em Phys.
  Rev. Lett.\/} {\bf 103} 246601

\bibitem{chen2010gate}
Chen J, Qin H, Yang F, Liu J, Guan T, Qu F, Zhang G, Shi J, Xie X, Yang C {\em
  et~al.\/} 2010 {\em Phys. Rev. Lett.\/} {\bf 105} 176602

\bibitem{checkelsky2011bulk}
Checkelsky J~G, Hor Y~S, Cava R~J and Ong N 2011 {\em Phys. Rev. Lett.\/} {\bf
  106} 196801

\bibitem{he2011impurity}
He H~T {\em et~al.\/} 2011 {\em Phys. Rev. Lett.\/} {\bf 106} 166805

\bibitem{liu2012crossover}
Liu M, Zhang J, Chang C~Z, Zhang Z, Feng X, Li K, He K, Wang L~l, Chen X, Dai X
  {\em et~al.\/} 2012 {\em Phys. Rev. Lett.\/} {\bf 108} 036805

\bibitem{zhang2012interplay}
Zhang D, Richardella A, Rench D~W, Xu S~Y, Kandala A, Flanagan T~C, Beidenkopf
  H, Yeats A~L, Buckley B~B, Klimov P~V {\em et~al.\/} 2012 {\em Phys. Rev.
  B\/} {\bf 86} 205127

\bibitem{yang2013}
Yang Q~I {\em et~al.\/} 2013 {\em Phys. Rev. B\/} {\bf 88} 081407

\bibitem{lang2013competing}
Lang M, He L, Kou X, Upadhyaya P, Fan Y, Chu H, Jiang Y, Bardarson J~H, Jiang
  W, Choi E~S {\em et~al.\/} 2013 {\em Nano Lett.\/} {\bf 13} 48--53

\bibitem{lu2011competition}
Lu H~Z, Shi J and Shen S~Q 2011 {\em Phys. Rev. Lett.\/} {\bf 107} 076801

\bibitem{lu2010massive}
Lu H~Z, Shan W~Y, Yao W, Niu Q and Shen S~Q 2010 {\em Phys. Rev. B\/} {\bf 81}
  115407

\bibitem{steinberg2011electrically}
Steinberg H, Lalo{\"e} J~B, Fatemi V, Moodera J~S and Jarillo-Herrero P 2011
  {\em Phys. Rev. B\/} {\bf 84} 233101

\bibitem{lin2013parallel}
Lin C, He X, Liao J, Wang X, Sacksteder~IV V, Yang W, Guan T, Zhang Q, Gu L,
  Zhang G {\em et~al.\/} 2013 {\em Phys. Rev. B\/} {\bf 88} 041307

\bibitem{shan2010effective}
Shan W~Y, Lu H~Z and Shen S~Q 2010 {\em New J. Phys.\/} {\bf 12} 043048

\bibitem{zhang2010crossover}
Zhang Y, He K, Chang C~Z, Song C~L, Wang L~L, Chen X, Jia J~F, Fang Z, Dai X,
  Shan W~Y {\em et~al.\/} 2010 {\em Nat. Phys.\/} {\bf 6} 584--588

\bibitem{li2017dirac}
Li M, Song Q, Zhao W, Garlow J~A, Liu T~H, Wu L, Zhu Y, Moodera J~S, Chan M~H,
  Chen G {\em et~al.\/} 2017 {\em Phys. Rev. B\/} {\bf 96} 201301

\bibitem{eremeev2013}
Eremeev S~V {\em et~al.\/} 2013 {\em Phys. Rev. B\/} {\bf 88} 144430

\bibitem{mathimalar2020signature}
Mathimalar S, Sasmal S, Bhardwaj A, Abhaya S, Pothala R, Chaudhary S, Satpati B
  and Raman K~V 2020 {\em npj Quantum Mater.\/} {\bf 5} 1--6

\bibitem{tkachov2011weak}
Tkachov G and Hankiewicz E 2011 {\em Phys. Rev. B\/} {\bf 84} 035444

\bibitem{stephen2020weak}
Stephen G~M, Vail O~A, Lu J, Beck W~A, Taylor P~J and Friedman A~L 2020 {\em
  Sci. Rep.\/} {\bf 10} 1--7

\bibitem{park2018disorder}
Park H, Chae J, Jeong K, Choi H, Jeong J, Kim D and Cho M~H 2018 {\em Phys.
  Rev. B\/} {\bf 98} 045411

\bibitem{lu2011weak}
Lu H~Z and Shen S~Q 2011 {\em Phys. Rev. B\/} {\bf 84} 125138

\bibitem{lu2014finite}
Lu H~Z and Shen S~Q 2014 {\em Phys. Rev. Lett.\/} {\bf 112} 146601

\bibitem{zhang2012weak}
Zhang H, Yu H, Bao D, Li S, Wang C and Yang G 2012 {\em Phys. Rev. B\/} {\bf
  86} 075102

\bibitem{liu2011electron}
Liu M, Chang C~Z, Zhang Z, Zhang Y, Ruan W, He K, Wang L~l, Chen X, Jia J~F,
  Zhang S~C {\em et~al.\/} 2011 {\em Phys. Rev. B\/} {\bf 83} 165440

\bibitem{chen2011tunable}
Chen J, He X, Wu K, Ji Z, Lu L, Shi J, Smet J and Li Y 2011 {\em Phys. Rev.
  B\/} {\bf 83} 241304

\bibitem{wang2011}
Wang J {\em et~al.\/} 2011 {\em Phys. Rev. B\/} {\bf 83} 245438

\bibitem{lee1985disordered}
Lee P~A and Ramakrishnan T 1985 {\em Rev. Mod. Phys.\/} {\bf 57} 287

\bibitem{wolos2016g}
Wolos A, Szyszko S, Drabinska A, Kaminska M, Strzelecka S, Hruban A, Materna A,
  Piersa M, Borysiuk J, Sobczak K {\em et~al.\/} 2016 {\em Phys. Rev. B\/} {\bf
  93} 155114

\bibitem{kohler1975g}
K{\"o}hler H and W{\"o}chner E 1975 {\em Phys Status Solidi B\/} {\bf 67}
  665--675

\bibitem{analytis2010two}
Analytis J~G, McDonald R~D, Riggs S~C, Chu J~H, Boebinger G and Fisher I~R 2010
  {\em Nat. Phys.\/} {\bf 6} 960--964

\bibitem{taskin2011berry}
Taskin A and Ando Y 2011 {\em Phys. Rev. B\/} {\bf 84} 035301

\bibitem{takagaki2012weak}
Takagaki Y, Jenichen B, Jahn U, Ramsteiner M and Friedland K~J 2012 {\em Phys.
  Rev. B\/} {\bf 85} 115314

\bibitem{roy2013two}
Roy A, Guchhait S, Sonde S, Dey R, Pramanik T, Rai A, Movva H~C, Colombo L and
  Banerjee S~K 2013 {\em Appl. Phys. Lett.\/} {\bf 102} 163118

\bibitem{bansal2011epitaxial}
Bansal N, Kim Y~S, Edrey E, Brahlek M, Horibe Y, Iida K, Tanimura M, Li G~H,
  Feng T, Lee H~D {\em et~al.\/} 2011 {\em Thin Solid Films\/} {\bf 520}
  224--229

\bibitem{bansal2014}
Bansal N {\em et~al.\/} 2014 {\em Appl. Phys. Lett.\/} {\bf 104} 241606

\bibitem{chen2014molecular}
Chen Z, Garcia T~A, De~Jesus J, Zhao L, Deng H, Secor J, Begliarbekov M,
  Krusin-Elbaum L and Tamargo M~C 2014 {\em J. Electron. Mater.\/} {\bf 43}
  909--913

\bibitem{zhang2011raman}
Zhang J, Peng Z, Soni A, Zhao Y, Xiong Y, Peng B, Wang J, Dresselhaus M~S and
  Xiong Q 2011 {\em Nano Lett.\/} {\bf 11} 2407--2414

\bibitem{jerng2013ordered}
Jerng S~K, Joo K, Kim Y, Yoon S~M, Lee J~H, Kim M, Kim J~S, Yoon E, Chun S~H
  and Kim Y~S 2013 {\em Nanoscale\/} {\bf 5} 10618--10622

\bibitem{liu2015gate}
Liu Y, Chong C, Jheng J, Huang S, Huang J, Li Z, Qiu H, Huang S and Marchenkov
  V 2015 {\em Appl. Phys. Lett.\/} {\bf 107} 012106

\bibitem{brahlek2014}
Brahlek M, Koirala N, Salehi M, Bansal N and Oh S 2014 {\em Phys. Rev. Lett.\/}
  {\bf 113} 026801

\bibitem{brahlek2015}
Brahlek M, Koirala N, Bansal N and Oh S 2015 {\em Solid State Comm.\/} {\bf
  215} 54--62

\bibitem{hikami1980}
Hikami S, Larkin A~I and Nagaoka Y 1980 {\em Prog. Theor. Phys.\/} {\bf 63}
  707--710

\bibitem{ioffe1960}
Ioffe A and Regel A 1960 {\em Prog. Semicond.\/} {\bf 4} 237

\bibitem{altshuler1979zero}
Altshuler B and Aronov A 1979 {\em Solid State Commun\/} {\bf 30} 115--117

\bibitem{lin2002recent}
Lin J~J and Bird J 2002 {\em J. Phys. Condens. Matter\/} {\bf 14} R501

\bibitem{bansal2012thickness}
Bansal N, Kim Y~S, Brahlek M, Edrey E and Oh S 2012 {\em Phys. Rev. Lett.\/}
  {\bf 109} 116804

\bibitem{kim2011thickness}
Kim Y~S, Brahlek M, Bansal N, Edrey E, Kapilevich G~A, Iida K, Tanimura M,
  Horibe Y, Cheong S~W and Oh S 2011 {\em Phys. Rev. B\/} {\bf 84} 073109

\bibitem{mathur2001random}
Mathur H and Baranger H~U 2001 {\em Phys. Rev. B\/} {\bf 64} 235325

\bibitem{olsen1962electron}
Olsen J~L 1962 {\em Interscience Publishers, New York\/}

\bibitem{ghaemi2010plane}
Ghaemi P, Mong R~S and Moore J~E 2010 {\em Phys. Rev. Lett.\/} {\bf 105} 166603

\bibitem{kim2013coherent}
Kim D, Syers P, Butch N~P, Paglione J and Fuhrer M~S 2013 {\em Nat. Commun.\/}
  {\bf 4} 2040

\bibitem{tkavc2019influence}
Tk{\'a}{\v{c}} V, V{\`y}born{\`y} K, Komanick{\`y} V, Warmuth J, Michiardi M,
  Ngankeu A, Vondr{\'a}{\v{c}}ek M, Tarasenko R, Vali{\v{s}}ka M, Stetsovych V
  {\em et~al.\/} 2019 {\em Physical review letters\/} {\bf 123} 036406

\bibitem{zhang2010first}
Zhang W, Yu R, Zhang H~J, Dai X and Fang Z 2010 {\em New J. Phys.\/} {\bf 12}
  065013

\bibitem{li2019quantitative}
Li H, Wang H~W, Li Y, Zhang H, Zhang S, Pan X~C, Jia B, Song F and Wang J 2019
  {\em Nano Lett.\/} {\bf 19} 2450--2455

\bibitem{fukuyama1982effects}
Fukuyama H 1982 {\em J. Phys. Soc. Japan\/} {\bf 51} 1105--1110

\bibitem{markiewicz1984localization}
Markiewicz R and Rollins C 1984 {\em Phys. Rev. B\/} {\bf 29} 735

\end{thebibliography}
\providecommand{\noopsort}[1]{}\providecommand{\singleletter}[1]{#1}
\providecommand{\newblock}{}

\end{document}